\documentstyle[psfig,tighten,preprint,aps,amsmath]{revtex}

\author{Franco Bagnoli\thanks{%
	also INFN and INFM, sezione di Firenze, Largo E. Fermi, 2 I-50125 Firenze,
	Italy}} 
\address{
	Dipartimento di Matematica Applicata, \\ Universit\`a di Firenze,\\ via S.
	Marta 3, \\ I-50139 Firenze, Italy. }
\author{Pietro Li\`o} 
\address{Dipartimento di Biologia Animale e Genetica, \\ 
	Universit\`a di Firenze,\\
	Via Romana 17, \\ I-50125 Firenze, Italy}
\title{Selection, Mutations and Codon Usage in Bacterial Model}

\begin{document}
\maketitle

\begin{abstract}
We present a statistical model of bacterial evolution based on the coupling
between codon usage and tRNA abundance. Such a model interprets this aspect
of the evolutionary process as a balance between the codon homogenization
effect due to mutation process and the improvement of the translation phase
due to natural selection. We develop a thermodynamical description of the
asymptotic state of the model. The analysis of naturally occurring sequences
shows that the effect of natural selection on codon bias not only affects
genes whose products are largely required at maximal growth rate conditions
but also gene products that undergo rapid transient increases.
\end{abstract}

\vspace{1cm}

appeared as J. theor. Biol. \textbf{173}, 217 (1995).

\section{Introduction}

The evolution of the genetic material is determined by the interactions
among mutations, random drift and natural selection.

The mutation rate seems to vary both among and within genomes, being
affected by many factors, such as the chromosomal position (Sharp {\it et al.%
}, 1989), the G+C content (Wolfe, 1991), the nearest neighbor bases (Blake 
{\it et al.}, 1992), the different efficiency of the repair systems between
the lagging and the leading DNA strands during replication and transcription
(Veaute \& Fuchs, 1993). Thus the molecular clock seems to tick at different
rate for different DNA positions.

The natural selection acts as a driving force at virtually all levels of the
genetic information processing and biological organization: from the DNA
stability, replication and transcription to messenger RNA life span and
translation efficiency, to the correct functioning of the gene products in
the building up and propagation of a living organism. Although in principle
all these constraints could interact in a very complex way, it is indeed
fruitful to try to untangle the role of each element.

We have focused on the translation process as a major source of fitness for
the bacterial cell.

Because of the degeneracy of the genetic code and the historical process of
codon capture (Osawa \& Jukes, 1989), different codons (synonymous codons)\
specify for the same amino acid. The bias in the codon usage is generally
not random and it is very marked in some procaryotic and eucaryotic genes.
The codon choice is specie-specific (Grantham {\it et al.}, 1980), being
roughly the same, although with different intensities, in all the genes
within a genome. The codon bias is thought to be an effect of the
improvement of the efficiency of the translation process under natural
selection.

Several authors have stressed that synonymous codons in {\it E. coli},
preferentially used in highly expressed genes, are translated by the most
abundant iso-acceptors tRNA species and that the levels of tRNAs in
bacterial cells depend on the amino acid usage in proteins (Ikemura, 1981a).

A good correlation between codon usage and tRNAs population has also been
found in {\it Mycoplasma capricolum} (Yamao {\it et al.}, 1991), in yeast
(Ikemura, 1981b) (Ikemura, 1985) and in chloroplast genomes (Morton, 1993).

The importance of codon usage as a substrate for natural selection is well
proved by the coding strategy of the coli phage {\it T4}: the early
expressed genes of {\it T4} have a codon pattern close to that of {\it E.
coli} while the late expressed genes show a preference towards the codons
read by the tRNAs expressed by the genome of the phage itself (Cowe \&
Sharp, 1991).

The codon usage influences the elongation rate in the translation process,
being the pairing between codon and anticodon of the diffusing tRNA the
rate-limiting step compared with the peptide bond formation and
translocation steps (Gouy \& Grantham, 1980), but it cannot be directly
related to the protein expression rate. In fact many experiments have
demonstrated that other crucial steps affecting the protein production rate
are the selection of mRNA by a ribosome, the initiation phase and the
phenomenon of polypeptidyl-tRNA drop off (Menninger, 1978).

Moreover the dynamic of elongation phase seems to be more complex, depending
also by the differences in the energy of codon anticodon pairing (Grosjean
\& Fiers, 1982) (Thomas {\it et al.}, 1988); thus differences in elongation
rates have been found for codons read by the same tRNA species (Sorensen \&
Pedersen, 1991).

Different considerations should be drawn on the evolutionary role of either
the minor or the major codons. Some authors have suggested that selection of
some of the rarer tRNA species may be the rate-limiting step in protein
synthesis (Varenne {\it et al.}, 1984): they hypothesize that the rate of
polypeptide elongation could act as a regulatory mechanism in gene
expression when clusters of rare codons undergo translation, i.e. when the
required tRNA concentration is low and rate-limiting. This hypothesis has
been confirmed experimentally by manipulating the concentrations of some
tRNAs and creating rate-limiting conditions. Varenne (Goldman, 1982)
(Varenne {\it et al.}, 1984) demonstrated experimentally that the presence
of rare codons in highly expressed coding regions is associated with pauses
in the synthesis of proteins. Many different contexts have been tested with
similar results: Hoekema (Hoekema {\it et al.}, 1987) replaced some major
codons present in the first region of the {\it PGK1} gene of {\it %
Saccharomices cerevisiae} with synonymous minor codons; Kinnaird (Kinnaird 
{\it et al.}, 1991) generated a cluster of three rare codons in the {\it GDH}
gene in {\it Neurospora crassa}. Chen and Inouye (Chen \& Inouye, 1990) have
demonstrated that the introduction of a cluster of a rare codon, AGG, in 
{\it lacZ} sequence of {\it E. coli}, lowers the expression of the gene;
moreover this effect is inversely correlated with the distance between the
initiation site and the position of rare codons.

Models that takes into account the influence of minor codons in the
initiation region of mRNA for protein production rate have been presented
(Liljenstrom \& von Heijne, 1987) (Liljenstrom \& Bomberg, 1987).

On the contrary Sharp and Li {(Sharp \& Li, 1986) have hypothesized that the
pattern of synonymous codon usage in regulatory genes reflects primarily the
relaxation of natural selection. In fact, the absence of rare codons from
highly expressed genes may well result from negative selection, while in low
expressed genes, selection against rare codons is very weak and so they can
accumulate under the pressure of mutation. }

Remarkably different considerations should be drawn for the major codons,
i.e. codons read by the most abundant tRNAs. There is evidence that the
major proteins are translated faster than other proteins and that the
elongation rate at major codons is faster than that at other codons
(Emilsson \& Kurland, 1990a).

Some authors (Emilsson \& Kurland, 1990a) (Emilsson \& Kurland, 1990b) have
hypothesized that growth rate, in bacteria, depends on tRNA abundance. Such
relation results in the so called growth maximization strategy and it is
consistent with a major codon preference strategy, in which an optimal
subset of codons is thought to be used very frequently in highly expressed
genes, and their cognate tRNA species are supposed to increase considerably
at high growth rates conditions. This is in agreement with the fact that the
genes for the major tRNA species are located inside rRNA operons for a
coordinate expression.

The growth rate dependence on tRNA abundances and other aspects of the
evolution of genes and genomes in bacteria can be faced with statistical
mechanics tools. Let us consider an homogeneous population of bacterial
genomes, subjected only to synonymous mutations and natural selection. If we
study its evolution in the proximity of a steady dynamical state, as for
instance a balanced growth state, we can interpret the interaction between
mutations and natural selection as a competition between randomness and
order. In this spirit we develop a statistical model from which bacterial
evolution emerges as an optimization process under environmental constraints.

\section{The Model}

In order to study quantitatively the competition between mutation and
natural selection in naturally occurring sequences, we concentrate on the
process of the elongation phase. We suppose that the rate limiting step in
this phase is the relative abundance of charged tRNAs. Thus we assume that
the time required by a ribosome to process a codon is inversely proportional
to the abundance of the charged cognate tRNAs in its vicinity. Taking into
account the data from Gouy and Grantham (Gouy \& Grantham, 1980), the
process of tRNA diffusion in cell extract results much faster than all
others processes involved in our schematization, and thus we do not consider
the effects of spatial gradients.

Furthermore, we assume that the efficiency of aminoacyl-tRNA synthetases is
not rate limiting, i.e. that the concentration of the charged tRNAs does not
vary with the translation process regardless of the codon composition of
mRNAs. This approximation is not completely fulfilled in real bacteria. From
the same source above (Gouy \& Grantham, 1980) we obtain that in average
only 85\% of the tRNA pool is acylated, indicating that the reaction rate of
synthetase is not substrate limited.

In this approximation we do not take into account the finite size of a
ribosome, nor the queuing of ribosomes, thus disregarding the positional
effects of codon usage, as considered elsewhere (Liljenstrom \& Bomberg,
1987).

With this assumption, the translation time does not depend on the order of
codons in mRNA and can be calculated as the sum of the time required for
each codon. Finally, we do not consider the influence of fluctuations, i.e.
we develop a sort of ``mean field'' model for the translation.

We indicate the portion of mRNA that code for a protein (or, alternatively,
the whole mRNA) with a vector ${\bf c}$; $c_i${\ ($i=1,\ldots ,61$) being
the number of codons of type }$i$ and $L=\sum_ic_i$ the length of the coding
region. {We denote with }$a_k${\ the relative abundance of acylated tRNA of
type }$k$ ($k=1,\ldots ,n_{tRNA}$; $\sum_{k=1}^{n_{tRNA}}a_k=1$). For {\it %
E. coli }$n_{tRNA}=45$, see also the following section. The ``translational
efficiency'' of codons by tRNAs is represented by means of a matrix $R_{ik}$
of size $61\times n_{tRNA}$. Again from (Gouy \& Grantham, 1980), the
transpeptidation and translocation phases do not weight heavily on the tRNA
cycle, and disregarding them, the mean time required to translate the codon $%
i$ is inversely proportional to $\rho _i=\sum_{k=1}^{n_{tRNA}}R_{ik}a_k$.${\ 
}${The mean translation time per codon }$\tau $ (in arbitrary time units)\ {%
is thus given by }

\begin{equation}
\label{tau}\tau ({\bf c}{)}=\frac 1L\sum_{i=1}^{61}\frac{c_i}{\rho _i}. 
\end{equation}

For simplicity, the elements $R_{ik}$ can be taken to be either $0$, for a
tRNA $k$ that does not pair with a codon $i$, or $1$, disregarding the
differences in energies in the codon-anticodon pairing discussed in the
previous section.

In bacteria, cell division and DNA\ replication are coupled with cell growth
(Zyskind \& Smith, 1992) (Ageno, 1992). In balanced growth conditions the
average number of genomes, the average number of cells and the cell mass
increase exponentially with the same rate.

The cell growth largely depends on the production rate of the most abundant
biopolymers, among which ribosomal proteins constitute the larger fraction.
The time $\tau $ in formula (\ref{tau}), calculated for an ``average''
ribosomal protein, can be considered proportional to the mean duplication
time. We assume that the natural selection tends to lower this time. A
somewhat opposite effect is due to the mutation process, that tends to
randomize the codon sequence. Since silent mutations do not change the coded
aminoacid and thus they do not affect the composition of proteins, they are
neutral for the selection over the protein functionality. However, silent
mutation are not neutral for the calculation of the duplication time $\tau $%
, because of the presence of the abundance of cognate tRNAs in formula (\ref
{tau}).

Let us start from a\ simplified situation, in which two synonymous codons
(say codon $0$ and $1$) are only read by two tRNAs (with respectively
abundances $a_0$ and $a_1$), while all other codons are grouped together
(say codon $2$) and in average are read by tRNAs with abundance $a_2$ ($%
a_0+a_1+a_2=1$). The number of synonymous codons is $c_0+c_1=$ $l$. We
consider only mutations between $0$ and $1$ and vice versa. Since $\tau $
does not depend on the order of the codons in the string, the various
strains can be grouped together according with the number $j$ of one's in
the mRNA ($j=c_1$). For each group $j$ there are $g_j=\dbinom lj$ strains.

We rewrite the formula (\ref{tau}) for the time $\tau _j$ required to
duplicate a bacterium belonging to group $j$ as 
\begin{equation}
\label{tauj}
\begin{split}
\tau _j&=\left( \dfrac j{a_1}+ 
\dfrac{l-j}{a_0}+\dfrac{L-l}{a_2}\right) =rj+q+z;  \\ 
 r&=\dfrac{a_0-a_1}{a_0a_1};\quad q=\dfrac l{a_0};\qquad z=\dfrac{Kl}{a_2%
};\qquad K=\dfrac{L-l}l. 
\end{split}
\end{equation}

The quantity $K^{-1}$ is the relative abundance of the synonymous codons 0
and 1 in the coding region. For $l\ll L$, $K\rightarrow \infty $; for $%
l\simeq L$, $K\simeq 0$. Since selection only acts on 0 and 1 codons, $%
K^{-1} $ give an indication of the influence of selection on the gene.

In the unit time interval, the number of duplications $\nu _j$ of the mass $%
M_j$ of bacteria belonging to group $j$ is $\nu _j=\tau _j^{-1}$, and thus 
\begin{equation}
\label{Mj}\frac{dM_j}{dt}=\nu _jM_j; 
\end{equation}
where for simplicity we set all constants to $1$ and we neglect the time
dependence of $M_j$.

In the limit of very low mutation rate, we can assume that for each
generation there is at most only one synonymous mutation, that changes codon 
$1$ to $0$ or vice versa. Let us indicate with $\mu /l$ the rate of mutation
per codon ($\mu \leq 1$). The average fraction of bacteria in group $j$ that
undergo a mutation per unit of time is $\mu \nu _j\,$. For bacteria in group 
$j$, a synonymous mutation changes $j$ of one unit. Including mutations, and
working with the mass $m_j=M_j/g_j$ of a single strain in group $j$, formula
(\ref{Mj}) becomes 
\begin{equation}
\label{mj}\frac{dm_j}{dt}=(1-\mu )\nu _jm_j+\frac jl\mu \nu _{j-1}m_{j-1}+ 
\frac{l-j}l\mu \nu _{j+1}m_{j+1}; 
\end{equation}
for $0\leq j\leq l$, assuming that $\nu _{-1}=\nu _{l+1}=0$.

The total mass of the bacterial population is $M=\sum_{k=0}^lg_km_k$. We can
derive from eq. (\ref{mj}) the evolution equation for the distribution
probability $p_j=m_j/M$ of different strains in the total population. In a
natural environment the exponential growth periods are sporadic, generally
followed by starvation phases. We mimic this alternation by means of the
normalization of distribution, assuming that the strains corresponding to a
very low probability are those eliminated by natural selection.

Since mutations do not change the total mass of population, we have, summing
up over $j$ in eq. (\ref{mj}) 
\begin{equation}
\label{num}\dfrac{dM}{dt}=\sum_{k=0}^lg_k\nu _km_k=M\sum_{k=0}^lg_k\nu
_kp_k=M\bar \nu ;
\end{equation}
obtaining 
$$
\dfrac{dp_j}{dt}=\frac d{dt}\dfrac{m_j}M=\frac 1M\dfrac{dm_j}{dt}-\dfrac{m_j%
}{M^2}\dfrac{dM}{dt}; 
$$
and thus 
\begin{equation}
\label{pj}
\begin{split}
\dfrac{dp_j}{dt} & =  \left[ (1-\mu )\nu _j-\bar \nu \right] p_j+\dfrac
jl\mu \nu _{j-1}p_{j-1}+
\dfrac{l-j}l\mu \nu _{j+1}p_{j+1}; \\ 
\bar \nu  & =  \sum_{k=0}^lg_k\nu _kp_k.
\end{split}
\end{equation}

Before dealing with these equations from a mathematical point of view, let
us put some thermodynamical considerations. The scenario is reminiscent of
statistical mechanics systems, in which there is competition between order
(the energy function to be minimized, related to the average duplication
time) and the entropy, mutuated by the temperature.

Each strain in group $j$ contributes with $\tau _{j\,}$ to the mean
duplication time. Taking into account the analogy between the duplication
time of strains and the energy levels of a statistical system, in the
equilibrium (stationary) state we can tentatively apply the methods from
equilibrium statistical mechanics. The maximization of the ``entropy'' $%
S=-\sum_{k=0}^lg_kp_k\ln p_k$ under the constraints $\sum_{k=0}^lg_kp_k=1$%
(normalization of probability distribution) and $\sum_{k=0}^lg_kp_k\tau
_k=const$ (in order to select the most probable probability distribution
within those with the same fitness), gives the Boltzmann distribution
(Landau \& Lifshitz, 1958) 
$$
p_j=C\exp (-\beta \tau _j). 
$$
The Lagrange multiplier $\beta $ can be considered as the inverse of an
effective ``temperature'' $T$, that, intuitively, should be related to the
mutation rate $\mu $. In the stationary state, $\bar \nu $ is constant in
time, and from eq. (\ref{num}), $M$ (and thus $m_j$) grows exponentially.

Inserting the {\it Ansatz} 
$$
m_j=M_0\exp \left( \alpha t-\beta \tau _j\right)  
$$
($M_0$ is the total mass at time $t=0$) in the equation (\ref{mj}), and
using the relations (\ref{tauj}), we get for the asymptotic state 
\begin{equation}
\label{x}(1-\mu )\nu _j-\alpha +\frac{l-j}l\mu \nu _{j+1}x+\frac jl\mu \nu
_{j-1}\frac 1x=0,
\end{equation}
where $x=\exp (-\beta r)$. Approximating $\nu _{j+1}\simeq \nu _{j-1}\simeq
\nu _j$, with an error of order $r/\tau _j^2$, which is small for $l$ or $L$
large or $a_1-a_0$ small, eq. (\ref{x}) becomes a linear equation in $j$.
This equation holds independently of $j$ if 
\begin{equation}
\label{xalpha}
\begin{split}
x & =  \dfrac{-(1-\mu )lr+\sqrt{(1-\mu )^2l^2r^2+4\mu ^2q(q+lr)}}{2\mu
(q+lr)}, \\ 
\alpha  & =  \dfrac{\mu (1-x^2)}{lrx}.
\end{split}
\end{equation}
Using the relations (\ref{tauj}) we get 
$$
x=\dfrac{(1-\mu )(a_1-a_0)+\sqrt{(1-\mu )^2(a_1-a_0)^2+4\mu
^2a_0a_1(1+K\tilde a_0)(1+K\tilde a_1)}}{2\mu a_0(1+K\tilde a_1)} 
$$
where $\tilde a_i=a_i/a_2$, $i=0,1$.

Note that $\alpha =\bar \nu $. Numerical simulations of eq.(\ref{pj}) agree
very well with this solution, and show that it is the only stable solution
for almost all initial distributions. 

The asymptotic form of $M_j=g_jm_j$ for $K=0$, $l\rightarrow \infty $ and $%
1\ll j\ll l$ is%
$$
M_{j\,}=M_0\frac{2^l\sqrt{2}}{\sqrt{\pi l}}\exp \left( \alpha t-\frac{\left(
2j-l\right) ^2}{2l}-\beta \tau _j\right) . 
$$
It is remarkable that although the distribution of $p_j$ is always a growing
function of $j$ (for $a_1>a_0$), the distribution $M_j$ is bell shaped, due
to the contribution of the multiplicity factor $g_j$.

The most probable group $j_{\text{max}}$ corresponding to the maximum of $%
M_j $ is, using the Stirling approximation for $g_j$, 
$$
j_{\text{max}}=\dfrac{lx}{x+1}. 
$$

For $L\gg l$ ($K\rightarrow \infty $), $x=1$, $T\rightarrow \infty $, $%
\alpha =0$ and $j_{\text{max}}=l/2$. The distribution $p_j$ is flat because
in this case the selection has no effect. This is due to the effective
neutrality of mutations inside a group $j$.

Let us consider in the following the opposite case $L=l$ ($K=0$), that
represents a bacterium in which an essential protein is totally composed by
the single aminoacid coded by $0$ and $1$ codons. This rather unrealistic
case makes our model equivalent to a 1D kinetic Ising model (Kawasaki, 1972)
in an external field without interactions among spins. In this case we can
analize in more details the influence of selection.

We can express $\alpha $ and $\beta $ in term of $\varepsilon =a_1-a_0$:%
\begin{equation*}
\begin{split}
x&=  \dfrac{(1-\mu )\varepsilon +\sqrt{(1-\mu )^2\varepsilon ^2+\mu
^2(1-\varepsilon ^2)}}{\mu (1-\varepsilon )} \\  
\alpha &= 
\dfrac{\mu (x^2-1)(1-\varepsilon ^2)}{4\varepsilon lx} \\  
\beta &=\dfrac{1-\varepsilon ^2}{4\varepsilon }\ln (x) 
\end{split}
\end{equation*}
with $0\leq \varepsilon \leq 1$.

For $\varepsilon \rightarrow 0$ ($a_1\simeq a_0$, $r\rightarrow 0$) we have $%
T=\beta ^{-1}=4\mu +O(\varepsilon ^2)$ and $\alpha =1/2l$. This relationship
confirms the intuitive corrispondence between temperature and mutation rate,
at least for moderate differences in the tRNA abundances. For $\mu
\rightarrow 0$, we get $j_{\text{max}}=l$, corresponding to a distribution
dominated by the fitter strains. Increasing the mutation rate $\mu $, $j_{%
\text{max}}$ decreases; extrapolating to infinite value of $\mu $ (beyond
the validity of our model), we get $j_{\text{max}}=l/2$, and the influence
of selection is vanished by mutations.

For $\varepsilon \rightarrow 1$ ($a_1\gg a_{00}$, $r\rightarrow -\infty $), $%
T\rightarrow \infty $ and $\alpha =(1-\mu )/l$. In this limit only the
fittest group $j=j_{\text{max}}=l$ survives. In this case all mutations are
deleterious, and their effect is to lower the growth rate $\alpha $ of the
total mass of bacterial population.

At this stage of development, our model is too simplified to allow
quantitative comparison with experimental data. However, it can be used as
an interpretative tool for the understanding of the mutation/selection
dynamics.

\section{Pertinent DNA Sequences Analysis and Discussion}

In order to estimate the degree of optimization between codons and tRNA
species we have analyzed 1530 {\it Escherichia coli} coding regions.

The choice of analyzing mainly the {\it E. coli} coding sequences depends
not only on the large amount of data in literature and the large number of
genes that have been sequenced (about 35\% of the genome, at this date) but
on the fact that in {\it E. coli} genome the effect of a directional
mutational pressure towards an increase in G+C or A+T content is very little
(about 51\% G+C content), as compared with other organisms like {\it %
Micrococcus luteus} (74\% G+C content) or {\it Mycoplasma capricolum} (25 \%
G+C content) (Sueoka, 1993) (Osawa \& Jukes, 1988b). This pressure, probably
due to mutations in the DNA polymerase system, acts particularly in
shrinking the codon and anticodon sets: in {\it E. coli} there are 75 tRNA
genes, and 45 types of anticodons, while in {\it Micrococcus luteus} or {\it %
Mycoplasma capricolum} these numbers are much lower (29 and 33 tRNA genes,
resp.). Besides, the large effects of the directional mutational pressure on
genomes composition probably results in weakening other existing functional
constraints.

For each gene we have considered the simple relationship (\ref{tau}) between
codons at position $i$, ($c_i$) and tRNA molecule abundances ($\rho _i$). We
have considered equal translational efficiency for all codons.

The effective tRNA species abundances at different duplication rates are not
exactly known; little differences in the total tRNA abundances have been
proved experimentally among different strains of {\it E. coli} (Jakubowski,
1984), but the relative abundances do not seem to vary largely even among
different species of enterobacteria: the data by Ikemura show that there is
a good correspondence between the relative abundances of the tRNA species
between {\it E. coli} and {\it Salmonella typhimurium}. Thus we have
approached the problem performing the calculations in two ways:

a) considering the data on the relative abundances of the tRNA species
published by Ikemura ($\tau _{\text{I}}$) (Ikemura, 1981a) (Ikemura, 1981b)
(Ikemura, 1985) and Jakubowski (Jakubowski,1984).

b) using the data on tRNA gene dosage published by Komine ($\tau _{\text{K}}$%
) (Komine {\it et al.}, 1990).

Both these two implementations are affected by different approximations.

In case (a)\ the data determined by Ikemura are not exhaustive. We have
approximated the abundances of the minor tRNAs at the value 0.1 with respect
to the abundance of Leu-tRNA normalized at 1. In case (b) we have considered
the abundance of tRNA species to be just proportional to the number of tRNA
genes; this assumption could be not completely true because different tRNA
gene clusters are regulated by different promoters and because of
differences in aminoacyl-tRNA synthetases abundances and catalytic
activities.

In Fig. 1 we report the average time spent per codon $\tau _{\text{I}}$ for
1530 coding sequences, both for the correct reading frame and for the +1 and
+2 reading frames. The sequences are ordered according with the value of $%
\tau _{\text{I}}$ of the correct reading frame.

In Fig. 2 we report $\tau _{\text{K}}$ for the same sequences of Fig. 1.

The comparison of the two figures shows that there is a good agreement
between the values of $\tau _{\text{I}}$ and $\tau _{\text{K}}$, although
the data obtained using the distribution of tRNA abundances from Ikemura
show a more marked separation between the correct reading frame and the +1
and +2 frames. This indicates that there is a good correspondence between
tRNA abundances and number of copy of the related tRNA gene, i.e. the most
abundant tRNAs are expressed by triplicated or quadruplicated genes while
the less abundant tRNA species are expressed by genes present in a single
copy.

In all the analyzed coding regions, the correct reading frame always
presents the lower crossing times, although the differences are more marked
for the highly expressed sets: the mean translation time of the correct
reading frame for highly expressed coding regions (as ribosomal genes, see
Fig. 1) is lower than those calculated for non highly expressed coding
regions, while the value for the +1 and +2 frames are equal or sometimes
greater.

We have also compared our results with the Sharp and Li's Codon Adaptation
Index ($CAI$)\ of each sequence. This index does not depend directly on the
abundances of tRNA but it is a measure of the bias in codon usage (Sharp \&
Li, 1987) defined as follows

\begin{equation}
\label{CAI}CAI=\left( \prod_{k=1}^lw_k\right) ^{\frac 1l};\text{ }CAI\in
\left( 0,1\right) \text{ } 
\end{equation}
where the product of $w_k=w(i_k)$ is taken over the code under examination.
The quantities $w(i_k)$ for a codon $i$ at position $k$ is given by the
relative frequency of the codon $i$ with respect to the most used codon for
the same aminoacyd in a set of highly expressed genes. Note that in
principle is possible to sinthetize artificial sequences made only of major
codons with a $CAI$ value greater than that of the most abundant proteins.
This occours because this index neglect the influence of mutations. The same
misunderstanding occours with our quantity $\tau $.

In Fig. 3 we report the $CAI$ values for the same sequences of Fig. 1. The
agreement with both the $\tau _{\text{I}}$ and $\tau _{\text{K}}$ is rather
good.

The genes that show a highly biased codon usage have high values of $CAI$
(and low values of $\tau _{\text{I}}$) and belong to the translation and
transcription machinery, as for instance genes coding for ribosomal
proteins, initiation and elongation factors, heat shock or stringent
response proteins, genes involved in the core of intermediate metabolism and
genes coding for the most abundant membrane proteins.

Generally the genes whose products are proved to be required for maximal
growth rate show high value of $CAI$. We can consider for instance the{\it \ 
}$^{118}${\it IF2$\alpha \,$} and $^{71}${\it IF2$\beta $}(Sacerdot 
{\it et al.}, 1992) that operate in the initiation of the translation
process and probably affect the production of ribosomal RNA, and the product
of the{\it \ nusA} gene that is involved in the transcriptional process.
These three genes are clustered in the same operon, probably for a
co-regulation of translation and transcription. The apices indicate the
position of the sequences in the figures.

The DNA polymerases I and II are present in different concentrations; about
400 molecules of polymerase I per cell with respect to 17-100 molecules per
cell of polymerase II (Adams {\it et al.}, 1992). The $CAI$ values of the
two genes ($^{334}${\it polA} and $^{760}${\it polB}) reflect the
difference: 0.403 ($\tau _{\text{I}}$ = 2.29) versus 0.352 ($\tau _{\text{I}%
} $ = 2.89).

Since biosynthetic operons are quite constitutively expressed, except during
starvation conditions, the genes coding for the repressor proteins show $CAI$
values lower than repressor genes in catabolic operons.

In some cases genes coding for functional aggregates constituted by
structural proteins interacting in a precise stoichiometry ratio show values
of $CAI$ or $\tau _{\text{I}}$ close to this ratio. This relation is true
only for highly expressed genes; for example some ribosomal proteins, as{\it %
\ }$^1${\it L7} and $^{11}${\it L20}, are present as a tetramer in each
ribosome (Adams {\it et al.}, 1992): the $CAI$ value of the correspondent
genes are higher than those of other ribosomal genes; the same correlation
has been found for ATP synthetases genes that code for a large complex
constituted by many copies of different subunits: there is a good agreement
between the $CAI$ values of the genes and the stoichiometry of the subunits
in the complex. This correspondence almost vanishes for structural proteins
coded by low expressed genes as those coding for fimbriae, pili and flagella.

Generally, the enzymes that channel a metabolic pathway do not present the
same bias in codon usage, probably because natural selection acts primary on
the homogenization of the catalytic activity of the metabolon, however there
are also few interesting clues of the contrary.

The membrane lipids biosynthesis in {\it E. coli} involves the action of at
least 25 genes; many of these are clustered in the{\it \ fab} operon.
Although the regulation of this pathway is very complex and not yet
completely elucidated, with regard to both the total fatty acid content and
the phospholipids composition, there are some observations that indicate
that the activities of few enzymes (acetyl-CoA:ACP transacetylase, acetyl
CoA carboxylase) are rate limiting in vivo (Magnuson {\it et al.}, 1993). We
have found that the bias in codon usage of these genes is higher than that
of the genes coding for the other enzymes of the pathway. Since growth rate
depends also by membrane lipids production, the bias in the bottleneck of
the pathway probably increases the rate of production, while its effect on
the other genes is vanishingly small.

Other interesting examples regard the relatively high bias in codon usage in
genes like $^{24}${\it spoT}, $^{180}${\it htpr}, $^{280}${\it uspA}, $%
^{295} ${\it fis}, $^{130}$c{\it spA} and $^{175}${\it ackA}.

Since {\it spoT} encodes for a ppGpp pyrophosphohydrolase, it takes into
account for the degradation of ppGpp, the major inductor of the stringent
response, that accumulates during starvation periods.

Nutritional shift-up experiments, in which large amount of nutrients are
added to bacteria growing in a minimal medium, have revealed that there are
immediate changes in the rate of increase of cell mass while the cell
division rate continues for some time at pre-shift rate (Cooper, 1991).

Thus the competition among bacteria requires to exit quickly from the
starvation periods and to restart growing when conditions have changed; in
this sight the {\it spoT} gene ($CAI$= 0.59, $\tau _{\text{I}}$ =1.4 ) could
be a very important target for natural selection. Probably for the same
reason the {\it htpr} gene ($CAI$ = 0.553, $\tau _{\text{I}}$ = 1.95) has a
relatively highly biased codon usage: the encoded protein coordinates the
global response of the cell to heat shock events. In fact this gene codes
for a sigma-32 subunit that substitutes the sigma-70 in the core of RNA
polymerase making the new holoenzyme capable of activating the heat shock
promoters. Henceforth other heat shock genes show a biased codon usage as
for instance $^{485}${\it dnaJ} and $^{462}${\it groEL}. The intracellular
amount of the universal stress protein, coded by the gene {\it uspA} ($CAI$
= 0.52, $\tau _{\text{I}}$ = 2.19), greatly increases in the cases of
exhaustion of nutrients like carbon, nitrogen, phosphate, sulfate or
aminoacid (Nystrom \& Neidhardt, 1992).

We hypothesize that codon usage reflects not only the effects of selection
for high growth rate conditions but also selection for very fast, and
perhaps short time lasting, response to equally rapid environment changes. A
clue that agrees well with this hypothesis concerns the {\it fis} gene ($CAI$
= 0.53, $\tau _{\text{I}}$ = 2.21) whose product functions as a DNA binding
protein, involved in recombination reactions and in rRNA transcription. It
has been reported that when the availability of a rich medium allows the
bacterial cells to exit from the stationary phase, the amount of the {\it fis%
} gene product increases from less than 100 copies to over 50000 copies
before the first cell division. As the exponential growth goes on, the Fis
levels decrease considerably (Ball {\it et al.}, 1992). The gene {\it cspA} (%
$CAI$ = 0.808, $\tau _{\text{I}}$ = 1.81) codes for the major cold shock
gene (CS7.4) (Tanabe {\it et al.}, 1992); when the temperature shifts from
37 $^{\circ }C$ to 10 $^{\circ }C$ the level of the protein largely
increases and becomes about 13\% of the total proteins in the bacterial cell.

It is noteworthy that since acetyl phosphate seems to be a global regulator
of signal transduction in {\it E. coli}, also the gene {\it ackA} ($CAI$ =
0.66, $\tau _{\text{I}}$ = 1.9) that codes for a protein that synthesizes
acetyl phosphate from ATP and acetate presents a high value of $CAI$ .

There are genes with little if any differences in values of $\tau _{\text{I}%
} $ between the correct reading frame and the others. This is mostly true
for genes in plasmids (see Fig. 1). Since plasmids and conjugative
transposons play a key role in the interchange of drug resistance and
virulence traits among bacteria, and generally in the mobilization of the
genetic material, it has been proposed that the entire pool of genetic
information could be accessible to all members of the bacterial community,
considered as a single, heterogeneous organism. In this regard the emergence
of a bias in codon usage for a gene carried by a plasmid is treated unfairly
by natural selection if the rate of the genetic information interchange
among conjugative bacteria of different species is sufficiently high. In
fact because of the genetic flux, the plasmid could visit a wide range of
hosts, thus it could assay different molecular environments and therefore
the recombinant processes between homologous genes maintain sequence
homogeneity. Furthermore, an increased expression of a gene could be
attained also by increasing the number of copies of the plasmids .

The debate between neutralists and selectionists in evolutionary and
population genetics addresses the question of how the superposition and the
interactions between the different constraints acting on the coding regions
allow the tremendous amount of molecular genetic variation that natural
bacteria populations exhibit.

The comparison of the standard deviation of the $CAI$ index for thirteen
sequences of D-glyceraldehyde-3-phosphate dehydrogenase gene ($^{33}${\it gap%
} gene; average $CAI$ = 0.830, standard deviation 0.0025) from different
strains of {\it E. coli} sequenced by Selander ( Selander {\it et al.},
1991) and ten sequences of the alkaline phosphatase gene ($^{843}${\it phoA}
gene, $CAI$ = 0.344, standard deviation 0.0059) sequenced by Milkman (DuBose
\& Hartl, 1991), shows that genes with higher bias in codon usage present
lower levels of silent polymorphism.

This relation suggests that the more biased the codon usage of a gene among
a bacteria population, the less neutral the synonymous mutation.

Another fundamental question regards if and in what measure the bias in
codon usage participates in the control of gene expression. {\it E. coli}
and bacteria in general are governed by robust and plastic regulative
mechanisms and even if regulative circuitry are parsimonious in design, the
present day gene networks have a large tool-kit of regulative mechanisms,
both at the gene-protein and at the protein-protein interaction level.

Although the basic mechanisms act at the operon level, there are global
mechanisms that allow the cell to respond to challenges, as hunger or stress
situations, with the coordinate and concerted expression of a network of
operons. These mechanisms rely mainly in a cascade-like activation of
autogenously regulated stimulons or in the sigma subunit variability that
makes the polymerases capable of recognizing different types of promoters.

Within the operon there are mechanisms that differentiate the expression at
the single gene level, as for instance internal promoters, transcription
termination signals, ribosome binding sites with different efficiency, mRNA
degradation signals etc.

Although bias in codon usage has been proved to play a role in
differentiating the expression levels of the genes within the operon, since
these mechanisms act sequentially, the effect of codon bias on the
translation process probably depends largely on the optimization of the
other factors involved in the gene expression.

The amounts of the genes that present high values of $CAI$, for instance the
ribosomal and major outer membrane genes whose products are strongly
required with respect to other proteins for high duplication rates, are the
effective fitness bottleneck of the bacterial duplication machinery; thus
the growth maximization strategy allows a cell to duplicate faster because
it lowers the time needed for translation.

The natural selection acts so much stronger in optimizing the translation of
the very highly expressed genes, with respect the other genes, because
further increases in the translation efficiency of these latter genes
reflects smaller improvements in the fitness of the cell.

Considerations of this type have also been proposed and tested by Dykhuizen
and collaborators for {\it lac} operon in {\it E. coli} (Dean {\it et al.},
1986); they have demonstrated that the contribution to the fitness of the
proteins encoded by the {\it lac} operon is quite different, being the
bottleneck of the lactose metabolic pathway the flux of lactose into the
cell, governed by the concentration of the enzyme lac permease: thus the
increase of catalytic activity of the other enzymes coded by the lac operon
(for example beta-galactosidase) results in a very little increase of the
global fitness of the pathway.

This could be a general principle in gene network architecture, i.e. the
potentiality of increasing the global fitness of the pathway could not be
equally distributed among the genes belonging to a gene network.

Being generally accepted that the weaker the selective constraint, the wider
the random genetic drift (Li \&{\it \ }Graur, 1991), the more relaxed
situation of genes whose product are not required in large amount, allows
high genetic variability consisting mainly in synonymous substitutions.

\section{Conclusions}

We have modeled some aspects of the molecular evolution of bacteria genomes
by means of the coupling between codon distributions and tRNAs abundance and
the competitive effects of mutation and selection. Our model is consistent
with a thermodynamical interpretation of this process, and gives a
matematical support to the observed codon usage distribution.

The evolution of a non synchronized population of bacteria is schematically
drawn as alternated phases of exponential growth (feast) and selection
(famine), due either to starvation or to external reasons. In bacteria,
because of the limiting amounts of nutrients and the rapid fluctuation in
their availability in natural environment, the periods of growth are
sporadic. Indeed, during these sporadic periods, the number of bacterial
cells tends to increase exponentially and the competition between genotypes
could develop very hardly, because a light difference in the fitness could
result in a large difference in number and in the affirmation of the fitter
genotype.

Probably the two extreme conditions of feast and famine model large part of
the fitness function of the bacterial genomes.

Bias in codon usage seems to be an adaptive response not only to feast
periods but also to conditions of changing surroundings. In fact to survive
famine periods would require sophisticated regulative mechanisms leading to
a shrewd management of the resources as for instance the stringent response,
while the selection for rapid oscillating periods make the cells maximize
the sharp up-shift or down-shift production of some specific proteins.
According to this hypothesis, bias in codon usage has revealed an
interesting tool to investigate the fitness bottleneck in metabolic pathways
or in gene networks like stringent response.

\-\vspace{1 cm}\ 

\noindent {\bf Acknowledgments}

We thank Fiammetta Battaglia, Marcello Buiatti and Stefano Ruffo for helpful
discussions.

\newpage\ 

\section{References}

\noindent 
Adams, L., Knowler, J. T., Leader, D. P. (1992). {\it The biochemistry of
the nucleic acids }Chapman \& Hall, London.

\noindent 
Ageno, M (1992). {\it La macchina batterica} (Lombardo ed.) Roma.

\noindent 
Andersson, G.E. \& Kurland, C.G. (1990).{\it \ Microbiol. Rev}.{\bf \ 5, }%
198-210.

\noindent 
Ball, C.A., Osuna, R., Ferguson, K.C., Johnson, R. C. (1992).{\it \ J. of
Bact}. {\bf 174, }8043-8056.

\noindent 
Berry, A.J., Ajioka, J.W., Kreitman, M. (1991). {\it Genetics }{\bf 129, }%
1111-1117.

\noindent 
Bibb, M.J., Findlay, P. R., Johnson, M.W. (1984). {\it Gene} {\bf 30, }%
156-166.

\noindent 
Blake, R.D., Samuel, T.H., Nicholson-Tuell, J. (1992). {\it J. Mol. Evol}. 
{\bf 34, }189-200.

\noindent 
Chen, G.T. \& Inouye, M. (1990). {\it Nucleic Acids Res.} {\bf 18, }%
1465-1473.

\noindent 
Cooper, S. (1991). {\it Bacterial Growth and Division}. Academic Press, San
Diego, California.

\noindent 
Cowe, E. \& Sharp, P.M.(1991). {\it J. Mol. Evol.} {\bf 33, }13-22.

\noindent 
Dean, A.M., Dykhuizen, D.E., Hartl, D.L. (1986). {\it Genet. Res. }{\bf 48, }%
1-8.

\noindent 
DuBose, R. \& Hartl, D.L. (1991). In: {\it Evolution at the Molecular Leve}l
(Selander, R.K. {\it et al }ed.) Sunderland, Massachusetts.

\noindent 
Emilsson, V. \& Kurland, C.G. (1990). {\it EMBO J.} {\bf 13,} 4359-4366.

\noindent 
Emilsson, V. \& Kurland, C.G. (1990). {\it Microbiol. Rev}. {\bf 54,}
198-210.

\noindent 
Eyre-Walker, A. \& Bulmer, M. (1993). {\it Nucleic Acids Res}. {\bf 21,}
4599-4603.

\noindent 
Goldman, E. (1982).{\it \ J.Mol. Biol}. {\bf 158, }619-636.

\noindent 
Gouy, M. \& Grantham, R. (1980). {\it Febs Letters} {\bf 115, }151-155.

\noindent 
Grosjean, H. \& Fiers, W. (1982). {\it Gene} {\bf 18, }199-209.

\noindent 
Grantham, R., Gautier, C., Gouy, M., Mercier, R., Pave, A. (1980) {\it %
Nucleic Acids Res.} {\bf 8, }49-62.

\noindent 
Hoekema, A., Kastelein, R.A., Vasser, M., De Boer H.A. (1987). {\it Mol.
Cell. Biol.} {\bf 7, }2914-2924.

\noindent 
Huynen, M.A., Konings, D. A.M., Hogeweg, P. (1992).{\it \ J.Mol. Evol}. {\bf %
34, }280-291.

\noindent 
Ikemura, T. (1981). {\it J.Mol. Biol. }{\bf 146,} 1-21.

\noindent 
Ikemura, T. (1981). {\it J.Mol. Biol.} {\bf 151,} 389-409.

\noindent 
Ikemura, T. (1985). {\it Mol. Biol. and Evol.} {\bf 2, }13-34.

\noindent 
Jakubowski, H., and Goldman, E. (1984). {\it J. of Bact}. {\bf 158, }769-776.

\noindent 
Jukes, T.H., Osawa, S., Muto, A., Lehman, N.(1987). {\it Cold Spring Harbor
Symposia on Quantitative Biology.} {\bf LII, }769-776.

\noindent 
Jukes, T.H., Ozeki, H., Umesono, K. (1988). {\it Proc. Natl. Acad. Sci. Usa }%
{\bf 85,} 1124-1128.

\noindent 
Kano, A., Andachi,Y., Ohama, T. , Osawa, S. (1991). {\it J. Mol. biol}. {\bf %
221,} 387-401.

\noindent Kawasaki, K. (1972). In: {\it Phase Transitions and Critical
Phenomena} (Domb, C. and Green, M.S. eds.), Academic Press, London.

\noindent 
Kinnaird, J. H., Burns, P.A., Fincham, J.R.S. (1991). {\it J. Mol. Biol. }%
{\bf 221, }733-736.

\noindent 
Komine, Y., Adaki, T., Inokuchi, H., Ozeki, H. (1990). {\it J.Mol. Biol. } 
{\bf 212, }579-598.

\noindent 
Landau, L.D. \& Lifshitz, E.M. (1958). In: {\it Statistical Physics},
Pergamon Press{\it , }Oxford, UK.

\noindent 
Li, W.H. \& Graur, D. (1991). {\it Fundamentals of Molecular Evolution}
Sinauer Associates inc. Publishers, Sunderland Massachussets.

\noindent 
Liljenstrom, H. \& von Heijne, G. (1987).{\it \ J. Theor. Biol. }{\bf 124, }%
43-55.

\noindent 
Liljenstrom, H. \& Blomberg, C. (1987).{\it \ J. Theor. Biol. }{\bf 129, }%
41-56.

\noindent 
Magnuson, K., Jackowski, S., Rock, C.O., Cronan, J. E. (1993). {\it Microb.
Rev.} {\bf 57, }522-540.

\noindent 
Menninger, J.R. (1978). {\it J. Biol. Chem} {\bf 253, }6808-6813.

\noindent 
Morton, B.R. (1993). {\it J Mol. Evol. }{\bf 37, }273-280.

\noindent 
Neidhardt, F.C., Ingraham, J.L., Law, J. (1987). In: {\it Escherichia Coli
and Salmonella Typhimurium: Cellular and Molecular Biology}. (Neidhardt, F.C.%
{\it et al.}, ed.) American Society for Microbiology, Washington, DC.

\noindent 
Nystrom, T. \& Neidhardt, F. C. (1992). {\it Mol. Microb.} {\bf 21, }%
3186-3198.

\noindent 
Jukes, T.H., Osawa, S., Muto, A., Leiman, N. (1987). {\it Cold Spring Harbor
symposium on quantitative biology } {\bf LII, }769-776.

\noindent 
Osawa, S. \& Jukes, T.H. (1988). {\it TIG} {\bf 7, }191-197.

\noindent 
Osawa, S., Ohama, T., Yamao, F., Muto, A., Jukes, T., Ozeki, H., Umesono, K.
(1988). {\it Proc.Natl. Acad. Sci. Usa } {\bf 85, }1125-1128.

\noindent 
Osawa, S. \& Jukes, T.H. (1989).{\it \ J.Mol. Evol. }{\bf 28, }271-278.

\noindent 
Ohama, T., Muto, A., Osawa, S. (1990). {\it Nucleic Acids Res.} {\bf 18, }%
1565-1569.

\noindent 
Sacerdot, C., Vachon, G., Laalami, S., Morel-Deville, F. , Cenatiempo, Y.,
Grunberg-Manago, M. (1992).{\it \ J. Mol. Biol. \ }{\bf 225, }67-80.

\noindent 
Nelson, K.N., Whittam, T.S., Selander, R.K. (1991). {\it Proc. Natl. Acad.
Sci. Usa } {\bf 88, }6667-6671.

\noindent 
Sharp, P.M. \& Li, W. (1986). {\it Nucleic Acids Res}. {\bf 14, }7737-7749.

\noindent 
Sharp, P.M. \& Li, W. (1987). {\it Nucleic Acids Res}. {\bf 15, }1281-1295.

\noindent 
Sharp, P.M., Shields, D.C., Wolfe, K.H., Li, W. (1989). {\it Science {\bf 258%
}, }808-810.

\noindent 
Sorensen, M. A., Kurland, C. G., Pedersen, S. (1989).{\it \ J.Mol. Biol.} 
{\bf 207, }365-377.

\noindent 
Sorensen, M. A., Jense, K.F., Pedersen, S. (1990). In: {\it %
Post-transcriptional control of gene \ expression}. (McCarthy and Tuite ed.)
NATO ASI Series.

\noindent 
Sorensen, M. A. \& Pedersen, S. (1991). {\it J.Mol. Biol.} {\bf 222, }%
265-280.

\noindent 
Sueoka, N. (1993).{\it \ J Mol Evol} {\bf 37, }137-153.

\noindent 
Tanabe, H., Goldstein, J., Yang, M., Inouye, M. (1992).{\it J. of Bacter.} 
{\bf 174}, 3867-3873.

\noindent 
Thomas, L.K., Dix, D.B., Thompson, R.C. (1988). {\it Proc. Natl. Acad. Sci.
Usa} {\bf 85, } 4242-4246.

\noindent 
Varenne, S., Buc, J., Lloubes, R., Lazdunsky, C. (1984).{\it \ J.Mol. Biol} 
{\bf 180, }549-576.

\noindent 
Varenne, S., Buc, J., Lloubes, R., Lazdunsky, C. (1986). {\it J.Theor.Biol.} 
{\bf 120, }99-110.

\noindent 
Veaute, X. \& Fuchs, R. (1993). {\it Science} {\bf 261, }598-601.

\noindent 
Yamao, F. , Andachi, Y., Muto, A., Ikemura, T., Osawa, S. (1991). {\it %
Nucleic Acids Res}. {\ {\bf 22, }}6119-6122.

\noindent 
Wright, S. (1932).{\it \ Proceedings of the Sixth International Congress of
Genetics }{\bf 1, }356-366.

\noindent 
Wolfe, K. H. (1991).{\it \ J. Theor.Biol.} {\bf 149, }441-451.

\noindent 
Zyskind, J.W., \& Smith, D.W. (1992).{\it \ Cell} {\bf 69},{\bf \ }5-8.

\begin{figure}
\centerline{\psfig{figure=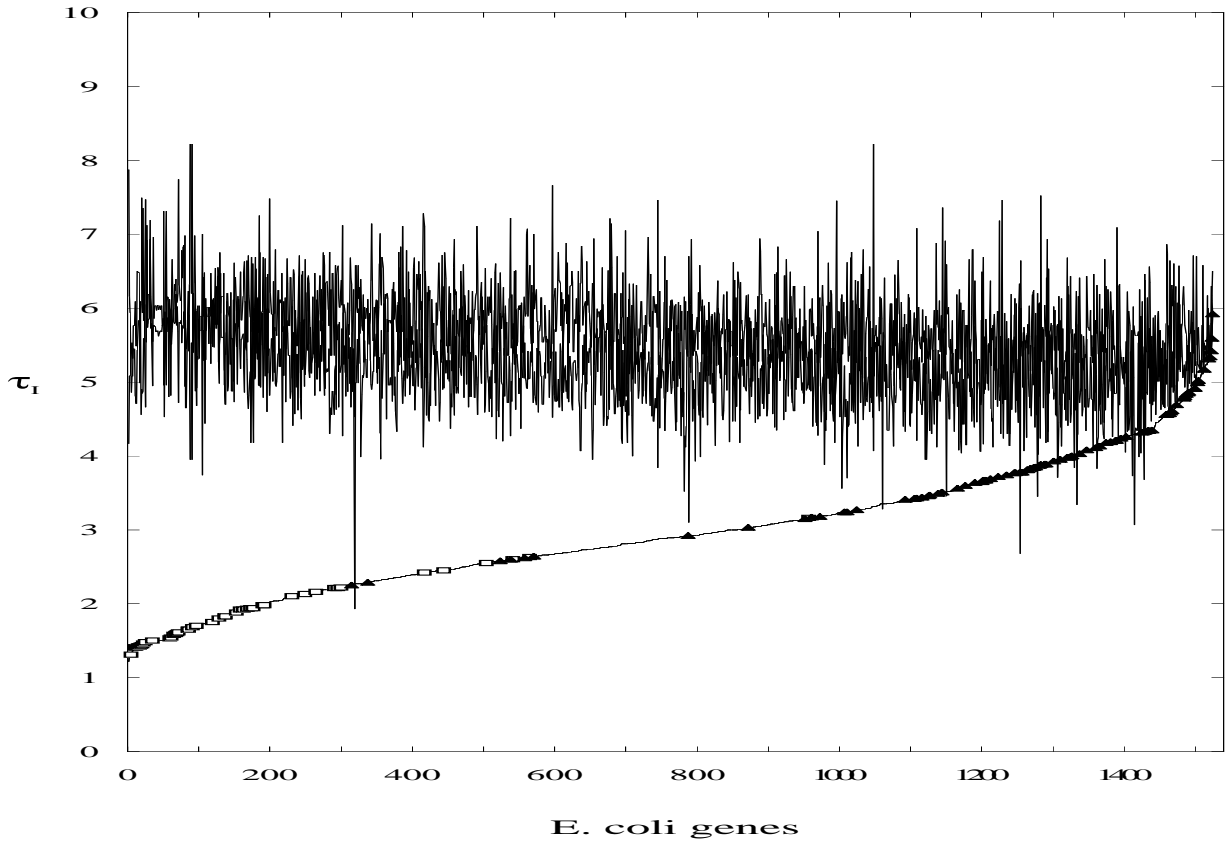,width=20cm,height=12cm}}
\caption{ Mean translation time $\tau _{\text{I}%
	} $ for 1530 {\it Escherichia coli} coding regions calculated using data on
	tRNA abundances published by Ikemura. The sequences are odered according to
	the values of $\tau _{\text{I}}$. The lower line shows $\tau _{\text{I}}$
	for the correct reading frame. The upper lines show the value of $\tau _{%
	\text{I}}$ for the +1 and +2 reading frames. The open squares correspond to
	ribosomal genes, the filled triangles correspond to genes carried by
	plasmids.}
\end{figure}

\begin{figure}
\centerline{\psfig{figure=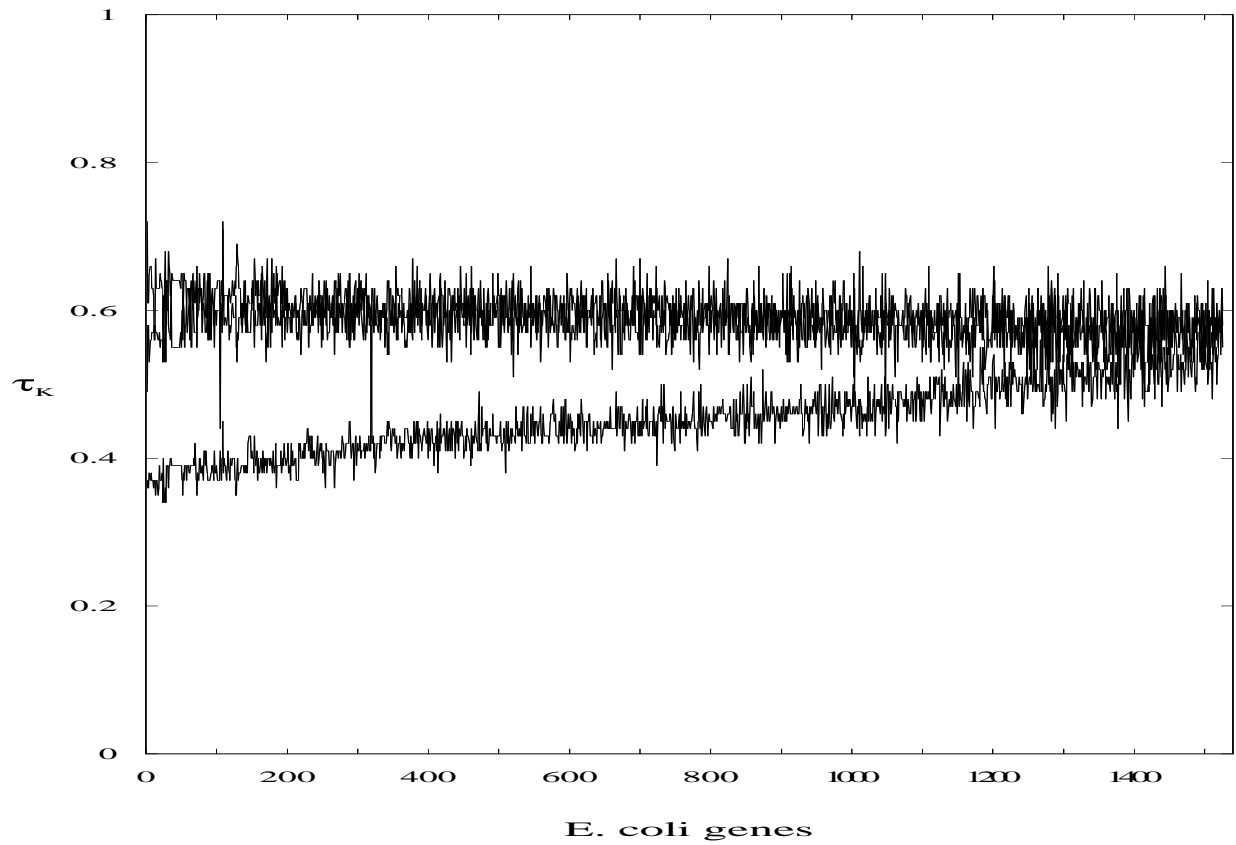,width=20cm,height=12cm}}
\caption{Mean translation time $\tau _{\text{K}}$ calculated using the data
	on tRNA gene dosage by Komine. The sequences are ordered as in Fig. 1. The
	lower line shows $\tau _{\text{K}}$ for the correct reading frame. The upper
	lines show the value of $\tau _{\text{K}}$ for the +1 and +2 reading
	frames.}
\end{figure}

\begin{figure}
\centerline{\psfig{figure=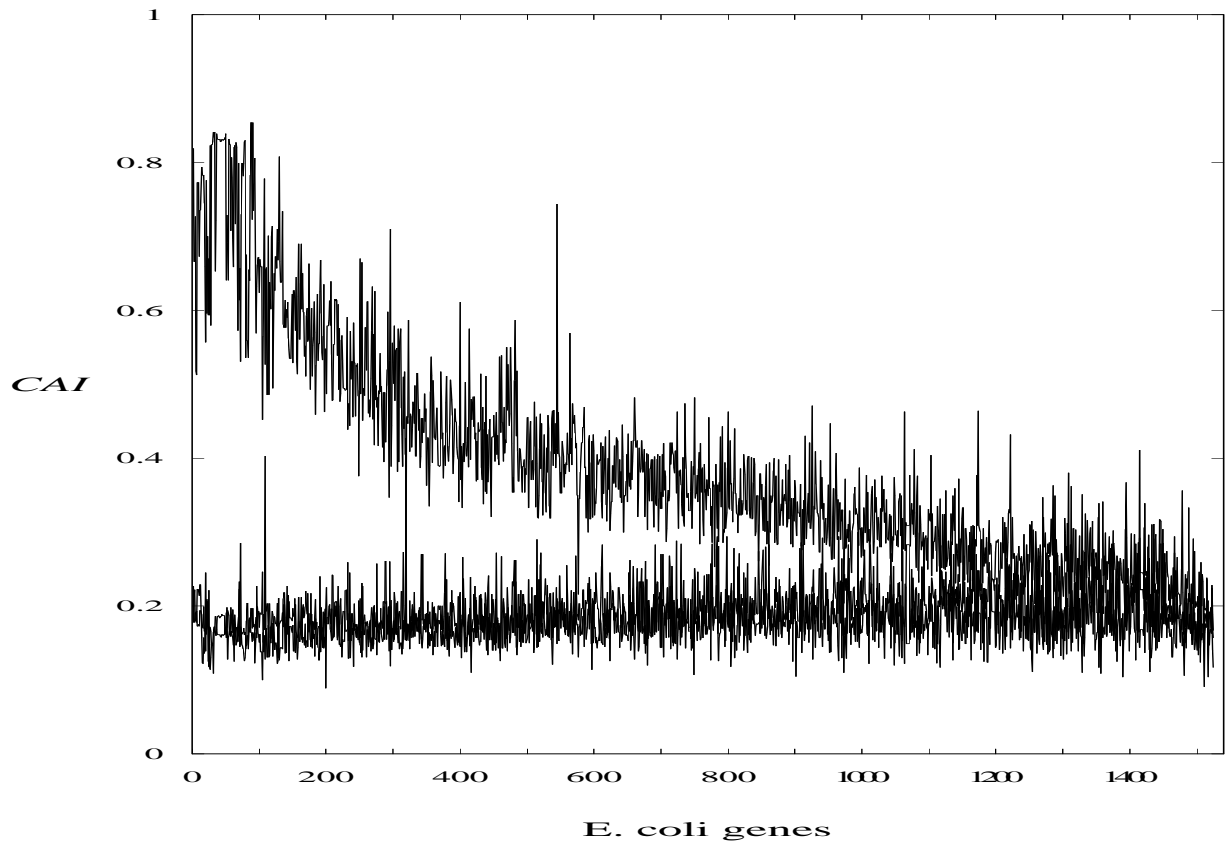,width=20cm,height=12cm}}
\caption{$CAI$\thinspace values of the coding sequences ordered as in Fig.
	1. The upper line shows $CAI$\thinspace values for the correct reading
	frame. The lower lines show $CAI$ values for the +1 and +2 reading frames.}
\end{figure}

\end{document}